\pdfoutput=1 
\documentclass[prb,singlecolumn,preprintnumbers,superscriptaddress,showpacs,preprint]{revtex4}
\usepackage{graphicx}
\usepackage{bm}
\usepackage{amsmath}
\usepackage{amssymb}
\usepackage[colorlinks,pdfstartview=FitH]{hyperref}
\hypersetup{linkcolor=blue,citecolor=blue,filecolor=black,urlcolor=blue}

\newcommand\be{\begin{eqnarray}}
\newcommand\ee{\end{eqnarray}}

\begin{document}


\author{Qiang Li}
\affiliation{Condensed Matter Physics and Materials Science Department, Brookhaven National Lab, Upton, New York 11973, USA}
\author{Dmitri E. Kharzeev}
\affiliation{Department of Physics, Brookhaven National Laboratory, Upton, New York 11973, USA}
\affiliation{Department of Physics and Astronomy, Stony Brook University, New York 11794-3800, USA}
\author{Cheng Zhang}
\affiliation{Condensed Matter Physics and Materials Science Department, Brookhaven National Lab, Upton, New York 11973, USA}
\author{Yuan Huang}
\affiliation{Center for Functional Nanomaterials, Brookhaven National Lab, Upton, New York 11973, USA}
\author{I. Pletikosi\'{c}}
\affiliation{Condensed Matter Physics and Materials Science Department, Brookhaven National Lab, Upton, New York 11973, USA}
\affiliation{Department of Physics, Princeton University, Princeton, NJ 08544, USA}
\author{A. V. Fedorov}
\affiliation{Advanced Light Source, Lawrence Berkeley National Laboratory, Berkeley, CA 94720, USA}
\author{R. D. Zhong}
\author{J. A. Schneeloch}
\affiliation{Condensed Matter Physics and Materials Science Department, Brookhaven National Lab, Upton, New York 11973, USA}
\author{G. D. Gu}
\affiliation{Condensed Matter Physics and Materials Science Department, Brookhaven National Lab, Upton, New York 11973, USA}
\author{T. Valla}
\affiliation{Condensed Matter Physics and Materials Science Department, Brookhaven National Lab, Upton, New York 11973, USA}

\title{Observation of the chiral magnetic effect in $\bf{ZrTe_5}$}

\begin{abstract}

The chiral magnetic effect is the generation of electric current induced by chirality imbalance in the presence of magnetic field. 
It is a macroscopic manifestation of the quantum anomaly\cite{Adler1969,Bell1969} in relativistic field theory of chiral fermions (massless spin $1/2$ particles with a definite projection of spin on momentum) -- a dramatic phenomenon arising from a collective motion of particles and antiparticles in the Dirac sea. The recent discovery\cite{Borisenko2014,Neupane2014,Liu2014} of Dirac semimetals with chiral quasi-particles opens a fascinating possibility to study this phenomenon in condensed matter experiments. Here we report on the first observation of chiral magnetic effect through the measurement of magneto-transport in zirconium pentatelluride, ZrTe$_5$. Our angle-resolved photoemission spectroscopy experiments show that this material\rq{}s electronic structure is consistent with a 3D Dirac semimetal. We observe a large negative magnetoresistance when magnetic field is parallel with the current.  The measured quadratic field dependence of the magnetoconductance is a clear indication of the chiral magnetic effect. The observed phenomenon stems from the effective transmutation of Dirac semimetal into a Weyl semimetal  induced by the parallel electric and magnetic fields that represent a topologically nontrivial gauge field background.

\end{abstract}

\maketitle

The recent discovery of three dimensional (3D) Dirac semimetals ${\rm Cd_3 As_2}$ \cite{Borisenko2014,Neupane2014} and ${\rm Na_3 Bi}$ \cite{Liu2014} enables experimental studies of the quantum dynamics of relativistic field theory in condensed matter systems. Relativistic theory of charged chiral fermions in three spatial dimensions possesses so-called chiral anomaly\cite{Adler1969,Bell1969} -- non-conservation of chiral charge induced by the external gauge fields with non-trivial topology, e.g. by parallel electric and magnetic fields. The existence of chiral quasi-particles in Dirac and Weyl semimetals opens the possibility to observe the effects of the chiral anomaly\cite{Nielsen1983}. Of particular interest is the chiral magnetic effect (CME)\cite{Kharzeev2008} -- the generation of electric current in an external magnetic field induced by the chirality imbalance, see\cite{Kharzeev2014} for a recent review and additional references.  

This phenomenon is currently under intense study in relativistic heavy ion collisions at Relativistic Heavy Ion Collider (RHIC) at BNL and at the Large Hadron Collider (LHC) at CERN, where it was predicted\cite{Kharzeev2006} to induce the fluctuations in hadron charge asymmetry with respect to the reaction plane. The experimental data from the STAR\cite{Abelev2009} Collaboration at RHIC and ALICE\cite{Abelev2013} Collaboration at LHC indicate the fluctuations consistent with the theory expectations.  Closely related phenomena are expected to play an important role in the Early Universe, possibly causing the generation of primordial magnetic fields \cite{Vilenkin1982,Frohlich2000,Joyce1997,Giovannini1998,Vachaspati2001}.
However, the interpretation in all these cases is under debate due to lack of control over the produced chirality imbalance.   

 The most prominent signature of the CME in Dirac systems in parallel electric and magnetic fields is the positive contribution to the conductivity that has a quadratic dependence on magnetic field \cite{Kharzeev2008,Son2013,Burkov2014}. This is because the CME current is proportional to the product of chirality imbalance and the magnetic field, and the chirality imbalance in Dirac systems is generated dynamically through the anomaly with a rate that is proportional to the product of electric and magnetic fields. As a result, the longitudinal magnetoresistance becomes negative\cite{Son2013,Burkov2014}. 
 
 Let us explain how this mechanism works in Dirac semimetals in more detail. 
 In the absence of external fields, each Dirac point initially contains left- and right-handed fermions with equal chemical potentials, $\mu_L = \mu_R = 0$.  If the energy degeneracy between the left- and right-handed fermions gets broken, we can parameterize it by introducing the chiral chemical potential $\mu_5 \equiv (\mu_R - \mu_L)/2$. The corresponding density of chiral charge is then given by \cite{Kharzeev2008}
\be\label{chir_density}
\rho_5 = \frac{\mu_5^3}{3 \pi^2 v^3} + \frac{\mu_5}{3 v^3}\ \left(T^2 + \frac{\mu^2}{\pi^2}\right),
\ee
where $\mu$ and $T$ are the chemical potential and the temperature, and $v$ is the Fermi velocity.

The chiral anomaly of quantum electrodynamics dictates that the parallel external electric and magnetic fields generate the chiral charge with the rate given by
\be
\frac{d \rho_5}{dt} = \frac{e^2}{4 \pi^2 \hbar^2 c} \vec{E}\cdot \vec{B} .
\ee
The left- and right-handed fermions in Dirac semimetals can however mix through chirality-changing scattering, and this process will deplete the amount of chiral charge that can be produced. Denoting the chirality-changing scattering time by $\tau_V$, we thus get the equation
\be\label{anomalyeq}
\frac{d \rho_5}{dt} = \frac{e^2}{4 \pi^2 \hbar^2 c} \vec{E}\cdot \vec{B} - \frac{\rho_5}{\tau_V} .
\ee
The solution of 
equation (\ref{anomalyeq}) at $t \gg \tau_V$ is
\be
\rho_5 = \frac{e^2}{4 \pi^2 \hbar^2 c} \vec{E}\cdot \vec{B} \ \tau_V .
\ee
According to (\ref{chir_density}), this leads to a non-zero chiral chemical potential $\mu_5$  (we assume that $\mu_5 \ll \mu, T$):
\be\label{chirpot}
\mu_5 = \frac{3}{4} \frac{v^3}{\pi^2} \frac{e^2}{\hbar^2 c}\ \frac{\vec{E}\cdot \vec{B}}{T^2 + \frac{\mu^2}{\pi^2}} \tau_V .
\ee
On the lowest Landau level, the spins of positive (negative) chiral fermions are parallel (anti-parallel) to the external magnetic field. Therefore, for a positive fermion to be right-handed (i.e., have a positive projection of spin on momentum) means moving along the magnetic field, and for a negative fermion -- moving against the magnetic field. The left-handed fermions will move in the opposite directions, so there will normally be no charge separation. However, if the densities of the right- and left-handed fermions are different, the currents of positive and negative charges do not compensate each other, and the system develops a net electric current -- this is the CME. The corresponding current can be computed\cite{Kharzeev2008} by field-theoretical method and is given by
\be\label{cme}
{\vec J}_{\rm CME} = \frac{e^2}{2 \pi^2}\ \mu_5\ \vec{B} .
\ee
The formulae (\ref{cme}) and (\ref{chirpot}) yield the final expression for the CME current: 
\be\label{cme_cur}
J_{\rm CME}^i = \frac{e^2}{\pi \hbar} \ \frac{3}{8} \frac{e^2}{\hbar c} \ \frac{v^3}{\pi^3}\ \frac{\tau_V}{T^2 + \frac{\mu^2}{\pi^2}}\ B^i B^k E^k  \equiv \sigma^{ik}_{\rm CME} \ E^k.
\ee
We see that the CME is described by the conductivity tensor $\sigma^{ik}_{\rm CME} \sim B^i B^k$. When the electric and magnetic fields are parallel, 
the CME conductivity is 
\be\label{cme_cond}
\sigma^{zz}_{\rm CME} = \frac{e^2}{\pi \hbar} \ \frac{3}{8} \frac{e^2}{\hbar c} \ \frac{v^3}{\pi^3}\ \frac{\tau_V}{T^2 + \frac{\mu^2}{\pi^2}}\ B^2.
\ee 
Since the CME current is directed along the electric field, it will affect the measured conductivity as the total current will be the sum of the Ohmic and CME ones:
\be\label{ohm}
J = J_{\rm Ohm} + J_{\rm CME} = (\sigma_{\rm Ohm} + \sigma_{\rm CME}) \ E,
\ee
where $\sigma_{\rm CME} \equiv \sigma^{zz}_{\rm CME}$. 
If the electric and magnetic fields are parallel ($\theta =0$), there is no conventional contribution to magnetoresistance induced by the Lorentz force. The magnetoconductance (\ref{cme_cond}) has a characteristic quadratic dependence on magnetic field.
It is precisely this contribution to magnetoconductance with a quadratic dependence on magnetic field that we have unambiguously observed in ZrTe$_5$. 

ZrTe$_5$ is a layered material that crystallizes in the layered orthorhombic crystal structure, with prismatic ZrTe$_6$ chains running along the crystallographic $a$-axis and linked along the $c$-axis via zigzag chains of Te atoms to form two-dimensional (2D) layers, stacked along the $b$-axis into a crystal. This material has been known for its large thermoelectric power, resistivity anomaly \cite{OkadaShigeto2013} and a large positive magnetoresistance.\cite{Tritt1999} It shows a semi-metallic electronic structure with extremely small and light ellipsoidal Fermi surface(s), centered at the center ($\Gamma$ point) of the bulk Brillouin zone (BBZ). The calculations predict a small direct gap at $\Gamma$ ($\simeq 50$ meV), but previous transport studies show semi-metallic behavior, with quantum oscillations indicating a tiny but finite Fermi surface. \cite{Whangbo1982,McIlroy2004,Kamm1985,Weng2014} Quantum oscillations show that the effective mass in the chain direction ($m_a^{\star}\simeq0.03 m_e$) is comparable to that in a prototypical 3D Dirac semimetal, Cd$_3$As$_2$. \cite{L.P.HeX.C.HongJ.K.DongJ.PanZ.ZhangJ.Zhang2014,Kamm1985}

\begin{figure}[htbp]
\begin{center}
\includegraphics[width=16cm]{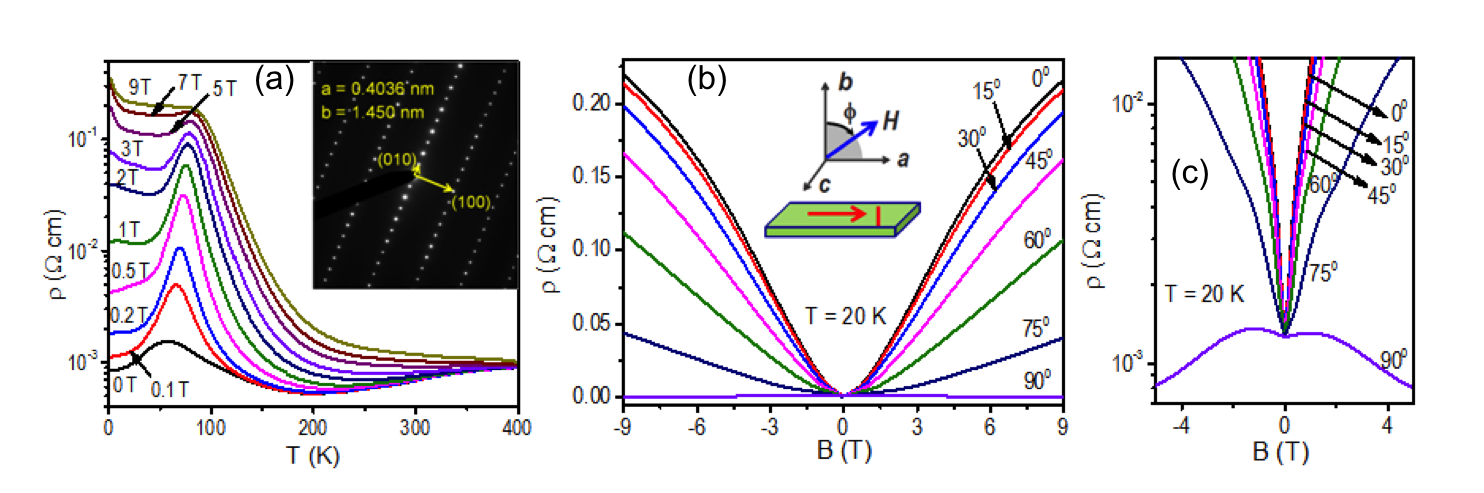}
\caption{Magnetoresistance in ZrTe$_5$.
(a) Temperature dependence of resistivity in ZrTe$_5$ in magnetic field perpendicular to the cleavage plane ($\vec{B}\parallel b$). The inset shows the electron diffraction from a single crystal looking down the (001) direction. 
(b) Magnetoresistance at 20 K for several angles of the applied field with respect to the current as depicted in the inset.
(c) The same data as in (b), plotted on the logarithmic scale, emphasizing the contrast between extremely large positive magnetoresistance for magnetic field perpendicular to current ($\vec{B}\parallel b$) and negative magnetoresistance for the field parallel to current ($\vec{B}\parallel a$).
}
\label{TR_1}
\end{center}
\end{figure}

Fig. \ref{TR_1}(a) shows temperature dependence of resitivity along the chain direction ($a$), in magnetic field perpendicular to the $a-c$ plane. The zero-field transport shows a characteristic peak in resistivity at $T\simeq60$ K, significantly lower than in earlier studies, probably due to a much lower impurity concentration in our samples. \cite{OkadaShigeto2013} In $\vec{B}\parallel b$, the peak shifts to higher temperatures and we observe a very large classical positive magnetoresistance in the whole temperature range, consistent with previous studies. \cite{Tritt1999}   

Panels (b-c) in Fig. \ref{TR_1} show the MR measured at 20 K for several angles of the applied magnetic field with respect to the current along the chain direction. The angle rotates from $b$- to $a$-axis, so that at $\phi=90^{\circ}$, the field is parallel to the current ($\vec{B}\parallel a$) - the 
so-called Lorentz force free configuration. When magnetic field is aligned along the $b$-axis ($\phi=0$), the MR is positive and quadratic  in low fields, and tends to saturate in high fields, consistent with a classical behavior. \cite{Pippard1989}
When magnetic field is rotated away from the $b$-axis, the positive MR drops with $\cos\phi$, as expected for the Lorentz force component. However, in the Lorentz force free configuration ($\vec{B}\parallel a$), we see a large negative MR, a clear indication of CME in this material.

\begin{figure}[htbp]
\begin{center}
\includegraphics[width=14cm]{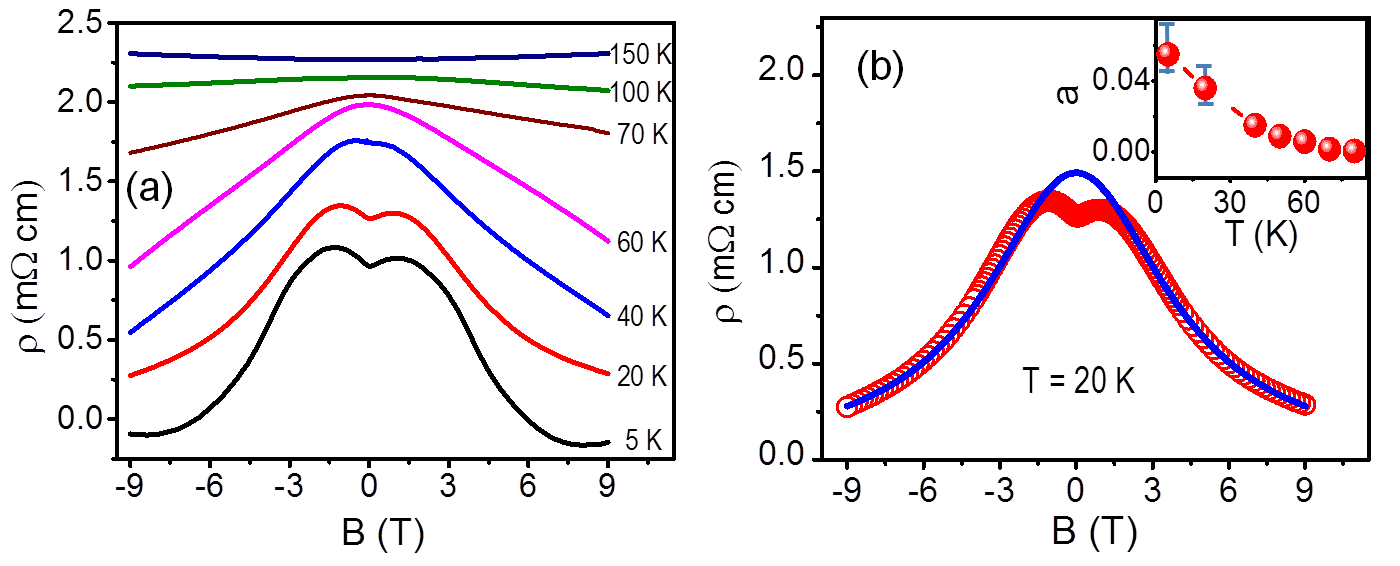}
\caption{Magnetoresistance in field parallel to current ($\vec{B}\parallel a$) in ZrTe$_5$.
(a) MR at various temperatures. For clarity, the resistivity curves were shifted by 1.5 m$\Omega$cm (150 K), 0.9 m$\Omega$cm (100 K), 0.2 m$\Omega$cm (70 K)  and $-0.2$ m$\Omega$cm (5 K).
(b) MR at 20K (red symbols) fitted with the CME curve (blue line); inset: temperature dependence of the fitting parameter $a(T)$ in units of ${\rm S}/({\rm cm}\ {\rm T}^2)$.
}
\label{TR_2}
\end{center}
\end{figure}
%
Fig. \ref{TR_2} shows the MR at various temperatures in a magnetic field parallel to the current. At elevated temperatures,  $T\geq 110$ K, the $\rho$ vs $B$ curves show a small upward curvature, a contribution from inevitable perpendicular field component due to an imperfect alignment between current and magnetic field. In fact, the small perpendicular field contribution to the observed resistivity can be fitted with a simple quadratic term (Supplementary materials, Fig. \ref{TR_S1}). This term is treated as a background and subtracted from the parallel field component for all MR curves recorded at $T\leq100$ K. 

A negative MR is observed for $T\leq100$ K, increasing in magnitude as temperature decreases. We found that the magnetic field dependence of the negative MR can be nicely fitted with the CME contribution to the electrical conductivity, given by  $\sigma_{CME}=\sigma_0+a(T)B^2$, where $\sigma_0$ represents the zero field conductivity. The fitting is illustrated in Fig. \ref{TR_2}(b) for $T=20$ K, with an excellent agreement  between the data and the CME fitting curve. At 4 Tesla, the CME conductivity is about the same as the zero-field conductivity. At 9T, the CME contribution increases by $\sim 400\%$, resulting in a negative MR that is much stronger than any conventional one reported at an equivalent magnetic field in a non-magnetic material.

At very low field, the data show a small cusp-like feature. The origin of this feature is not completely understood, but it probably indicates some form of anti-localization coming from the perpendicular ($\vec{B}\parallel b$) component. 
Inset in Fig. \ref{TR_2}(b) shows the temperature dependence of the fitting parameter $a(T)$, which decreases with temperature faster than $1/T$, again consistent with the CME.

\begin{figure}[htbp]
\begin{center}
\includegraphics[width=8cm]{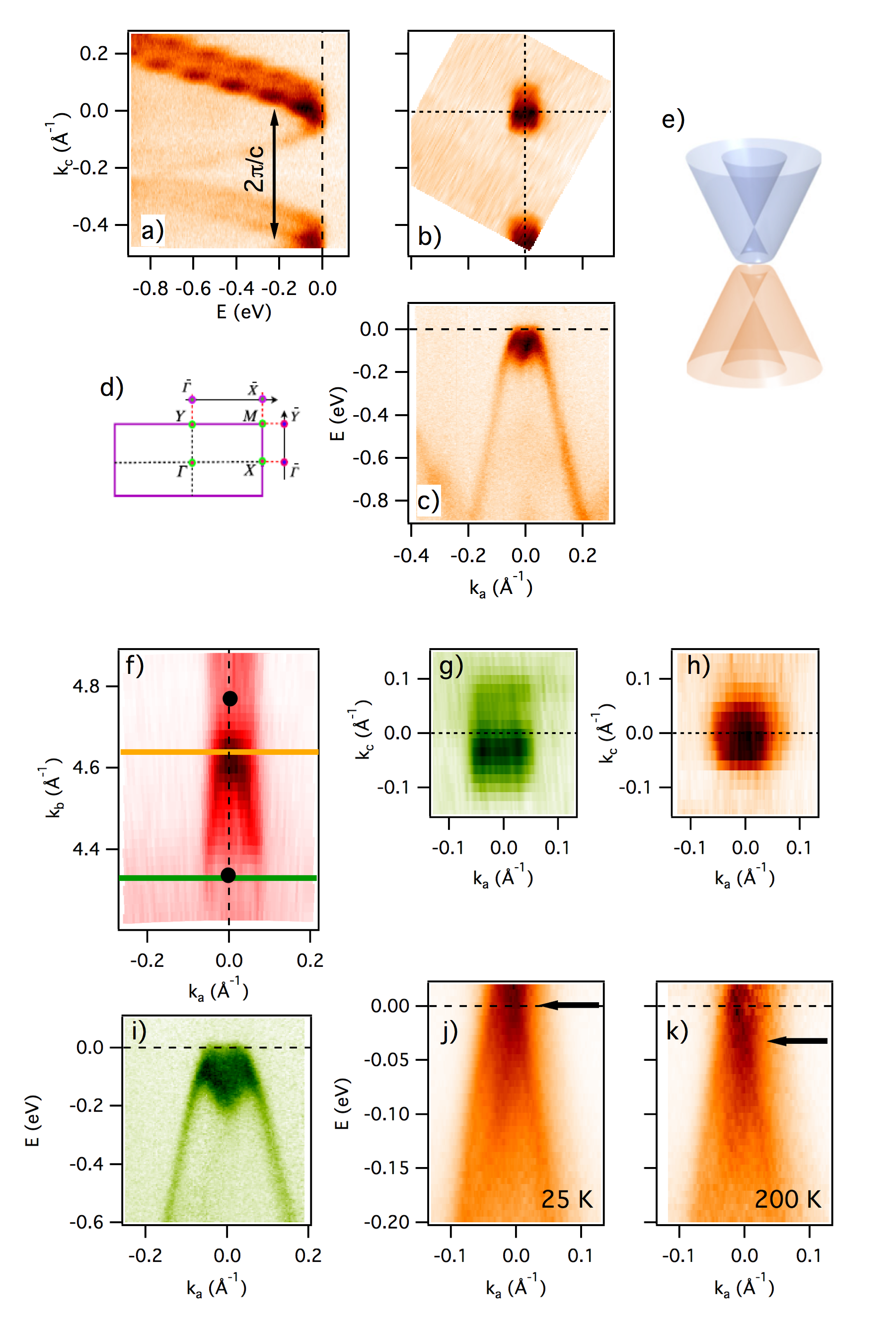}
\caption{Electronic structure of ZrTe$_5$. 
\textit{in-plane}: (a) Valence band dispersion along the $\bar\Gamma-\bar Y$ momentum line (perpendicular to the chain direction). The second BZ $\bar\Gamma$ point is visible at $k_c=-0.454$ \AA$^{-1}$.
(b) Constant energy contour of the ARPES intensity at $E=0$ (FS)
as a function of the in-plane momentum. 
(c) Valence band dispersion along the $\bar\Gamma-\bar X$ (chain direction). Spectra in (a-c) were recorded at $h\nu=60$ eV photon energy and $T=20$ K.
(d) Schematic view of the SBZ.
(e) Schematic view of the in-plane low-energy electronic structure.
\textit{out-of plane}: (f) Fermi surface contour as a function of in-plane momentum along the chain and momentum perpendicular to the surface. The solid circles represent zone centers of the 10$^{th}$ and 11$^{th}$ BBZ. Thick green and orange lines correspond to the two perpendicular momenta where the in-plane electronic structure was probed in panels (g) to (k).
(g) Fermi surface as a function of in-plane momenta, taken at $h\nu=60$ eV photon energy, corresponding to  $k_b\simeq 4.33$ \AA$^{-1}$ (green line in (f)) and 
(h) at $h\nu=21$ eV, ($k_b\simeq 2.92$ \AA$^{-1}$), equivalent to the $k_b$ in the 11$^{th}$ BZ marked by the orange line in (f).
(i) Dispersion along the chain direction at $k_c=0$, $k_b=4.33$ \AA$^{-1}$.
(j) - (k) Dispersion along the chain direction at $k_c=0$, $k_b=2.92$ \AA$^{-1}$ at $T\simeq25$ and 200 K. 
The spectra were divided by the corresponding Fermi distribution. The black arrows indicate the position of the Dirac point of the lower Dirac cone.
}
\label{ARPES}
\end{center}
\end{figure}

A necessary requirement for observation of the CME is that a material has a 3D Dirac semimetal-like (zero gap), or semiconductor-like (non-zero gap) electronic structure. Figure \ref{ARPES} shows angle-resolved photoemission spectroscopy (ARPES) data from a freshly cleaved ($a-c$ plane)
 ZrTe$_5$ sample. The states forming the Fermi surface (FS) disperse linearly over a large energy range, both along the chain direction (panel (c)) and perpendicular to it (panel (a)), indicating a Dirac-like dynamics of carriers for the in-plane propagation. The velocity, or the slope of dispersion, is very large in both the chain direction, $v_a\simeq 6.4$ eV\AA  ($\simeq c/300$), and perpendicular to it, $v_c\simeq4.5$ eV\AA.
 However, it is obvious that the in-plane electronic structure cannot be described by a single (anisotropic) cone, especially not at very low energies. The states are doubled - this is because the crystal contains two layers per unit cell - and the simplest description then would be that the bi-layer splitting creates two cones (bonding-antibonding), separated in energy by $\sim 300$ meV and possibly gapped by a small gap at the degeneracies, as schematically shown in panel (e). 

The low-energy electronic structure might be more complicated than that (see SOM, Fig. \ref{ARPES_S}), but due to a sizable quasiparticle scattering, $\Gamma_Q = \hbar/\tau_Q \simeq 100$ meV, the states appear too broad to completely resolve possible fine features in the low energy electronic structure, such as the existence of small gaps or changes in the velocity at low energies ($E<50$ meV). Nevertheless, the magnitude of chiral magnetic effect in ZrTe$_5$ is not particularly sensitive to the fine details in the electronic structure on the energy scale smaller that $\Gamma_Q$. On the other hand, for the existence of CME, it is crucial that the electronic structure of a material is 3D Dirac-like. Recent theoretical study has suggested that ZrTe$_5$ might be either weak or strong topological insulator (WTI or STI), with topological surface states present on all (STI), or just on the side (WTI) surfaces.\cite{Weng2014} It is therefore important to verify that the states seen in ARPES are bulk states and not topological surface states.

This is done in Figs. \ref{ARPES}(f-k) which show the dependence of the states forming the FS on the momentum $k_b$ perpendicular to the $a-c$ plane. Panel (f) shows a clear warping of the Fermi surface with $k_b$, indicating the bulk character of the measured states. It seems that in addition to the variation in the FS\rq{}s cross section area, its topology also varies with $k_b$: while close to the $k_b=0$ (or equivalent) the in-plane FS contour (panel (g)) seems to consists of four tiny ellipses, at $k_b\simeq\frac{3\pi}{4b}$ (and equivalent) it resembles two concentric circles (panel (h)).  
The true 3D Dirac part of the electronic structure then arises from the $k_a=k_c=0, k_b\simeq\frac{3\pi}{4b}$ region of the BBZ. The dispersion there seems to be linear in all three directions, at least for the lower-energy state, whose Dirac point sits at $E\simeq0$ at low temperatures, but shifts downward at higher temperatures, as illustrated in panels (j) and (k).
These spectra are consistent with the simple model shown in panel (e) and in conjunction with the dispersion shown in panel (f), they verify that ZrTe$_5$ is a 3D Dirac semimetal. We note that the temperature variation of the electronic structure (indicated in panels (j) and (k)) might be responsible for the large positive magnetoresistance seen in the perpendicular field, as the balance between the tiny hole- and electron-pockets varies with temperature with the possibility of perfect compensation at some temperatures.\cite{Pletikosic2014}

A finite experimental resolution and a significant quasiparticle scattering rate might mask a possible gap in the otherwise Dirac-like spectrum. The gap $\Delta$ will in general induce chirality-changing transitions with the rate $\Delta/\hbar$ -- this is because the corresponding term in the effective hamiltonian $\Delta ({\bar\Psi}_L \Psi_R + {\bar\Psi}_R \Psi_L)$ mixes the left and right components of the spinors -- but it does not prevent the CME. The ARPES data also indicate that  the quantum scattering rate determined from the broadening of quasiparticles is $\Gamma_Q = \hbar/\tau_Q \simeq 100$ meV. Chirality conservation in the quasi-particle scattering processes implies that the chirality-changing rate should be only a small fraction of the quantum scattering rate, $\Gamma_V = \hbar/\tau_V \ll \Gamma_Q$. It is therefore reasonable to assume that the measured $\Gamma_Q$ represents the absolute upper limit to the chirality-changing transition rate.

We note that the formula (\ref{ohm}) shows that in the regime when $\rho_{\rm Ohm} \gg \rho_{\rm CME}$, the measured magnetoresistivity $\rho$ directly provides the CME resistivity, $\rho \simeq \rho_{\rm CME}$. Therefore it is much easier to observe the CME in materials that have a relatively large zero-field resistivity. Note that the zero-field resistivity in ${\rm ZrTe_5}$ is  $\rho_{\rm Ohm} \simeq {\rm 1.2\ m\Omega\ cm}$ at $T = 20$ K, and according to our theoretical estimates the CME and Ohmic resistivities become equal, $\rho_{\rm Ohm} \simeq \rho_{\rm CME}$, at $B \simeq 3$ T. On the other hand, if a material had a much smaller resistivity, the CME observation would have been more difficult.

The present study can be extended to a broad range of materials as 3D Dirac semimetals often emerge at quantum transitions between normal and topological insulators, including topological crystalline insulators. Moreover, our experimental observation has important implications extending well beyond condensed matter physics.  



\section*{Acknowledgements}
We thank J. Misewich, P. Johnson, A. Abanov and G. Monteiro for discussions. 
This work was supported by the US Department of Energy, Office of
Basic Energy Sciences, contracts No. DE-AC02-98CH10886, No. DE-FG-
88ER40388 and ARO MURI program, grant W911NF-12-1-0461. 
ALS is operated by the US DOE under Contract No. DE-AC02-05CH11231. 

\section*{Author Contributions}
D.E.K. designed and, with Q.L. and T.V., directed the study, analyzed results, and wrote the manuscript. Q.L. and C.Z. performed the transport measurements and analyzed results. R.D.Z., J.A.S.and G.D.G. grew the crystals and performed x-ray diffraction experiments, Y.H. performed the SEM/TEM measurements and provided analysis. I.P., A.V.F and T.V. performed the ARPES measurements and analyzed results. All authors made contributions to writing the manuscript.

\section*{Author Information}
 The authors declare no competing financial interests. Readers are welcome to comment on the online version of the paper. Correspondence and requests for materials should be addressed to Q.L. (liqiang@bnl.gov), D.E.K (dmitri.kharzeev@stonybrook.edu) and T.V. (valla@bnl.gov).

\renewcommand{\thefigure}{S\arabic{figure}}
\setcounter{figure}{0}

\section*{METHODS}

{\underline{Crystal Growth:}}

Crystals of ZrTe$_5$ were grown by flux growth method using Te as flux.
High purity elements (99.99999\% Te and 99.9999\% Zr) were loaded into a double-walled quartz ampule and sealed under vacuum. The composition of the Zr-Te melt used for the ZrTe$_5$ growth was Zr$_{0.0025}$Te$_{0.9975}$. The materials were first melted at 900$^{\circ}$C in a box furnace and fully rocked to achieve a homogeneous mixture for 72 hours. The melt was then slowly cooled and rapidly heated (for re-melting of small size crystals) between 445 and 505$^{\circ}$C for 21 days. The largest single crystals were  $\sim1\times1\times20$ mm$^3$. Crystals were 
chemically and structurally characterized by powder X-ray diffraction, scanning electron microscopy with energy dispersive x-ray analysis, and transmission electron microscopy with electron diffraction.

{\underline{Transport Measurements:}}

The magnetoresistance of ZrTe$_5$ samples was measured using the 4-point probe in-line method in a Quantum Design Physical Property Measurement System (PPMS) equipped with 9 Tesla superconducting magnet. For crystal alignment with magnetic field, horizontal and vertical sample rotators were used with the angular resolution ~ 0.1 degree. Temperature dependent data were taken from 1.8 to 400 K, at various fields up to 9 T. 

\begin{figure}[htbp]
\begin{center}
\includegraphics[width=10cm]{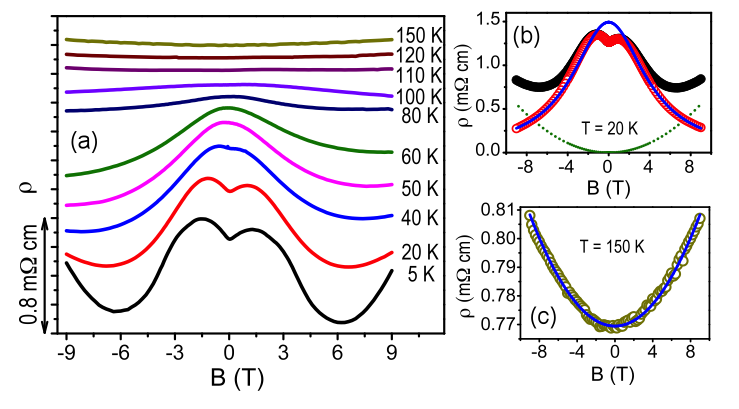}
\caption{Magnetic field dependence of the raw resistivity data at various temperatures for field parallel to current. For clarity, resistivity curves were shifted up or down as explained in the caption of Fig. \ref{TR_2}. (b) Raw MR data at 20K (solid black symbols) plotted together with the fitting curve based on the CME (solid blue line), simple parabolic background (dotted line), and the data after background subtraction (open red symbols). (c) Magnetic field-dependence of the raw resistivity data at 150K (open symbols) plotted together with the fitting curve (line) based on the simple quadratic approximation for small positive magnetoresistivity from the perpendicular field component due to a small misalignment. 
}
\label{TR_S1}
\end{center}
\end{figure}
%
The raw magnetoresistance data for the field parallel to the current, Fig. \ref{TR_S1}(a), show an upturn at higher fields coming from the perpendicular component due to an imperfect alignment of magnetic field and current. The perpendicular component, known from $\vec{B}\parallel b$ measurements, is then properly scaled and subtracted from the raw $\vec{B}\parallel a$ data as illustrated in Fig. \ref{TR_S1}(b-c). 

{\underline{ARPES Experiments:}}

The ARPES measurements were conducted using a Scienta SES2002 analyzer
at the U13 beam line of the National Synchrotron
Light Source at Brookhaven National Laboratory ($h\nu=21\,\mathrm{eV}$), and at the 12.0.1 beam
line of the Advanced Light Source at Lawrence Berkeley National Laboratory (38--78~eV) using a Scienta
SES100 analyzer. The total experimental resolution was $\simeq15$ meV in energy
and $\leq0.2^{\circ}$ in angle, in both experimental setups. The two-dimensional
Brillouin zone mapping was accomplished by sample rotation perpendicularly
to the analyzer slit, in steps of $1^{\circ}$ at $h\nu=21\,\mathrm{eV}$ and $0.5^{\circ}$ at higher photon energies. The samples were glued to the holder by a conductive epoxy resin and cleaved at the $a-c$ plane in ultrahigh vacuum ($p<10^{-8}~\mathrm{Pa}$)
just before the measurements. Sample cooling was provided through
contact with cryostats filled with liquid helium or liquid nitrogen.

\begin{figure}[htbp]
\begin{center}
\includegraphics[width=8cm]{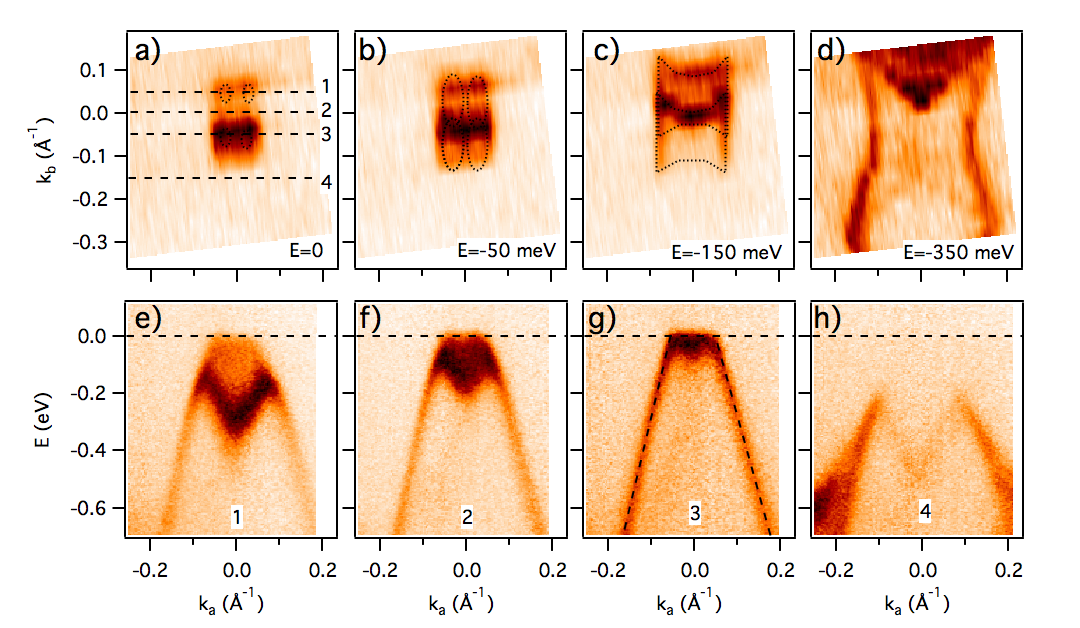}
\caption{ (a) - (d) Constant energy contours of the ZrTe$_5$ in-plane electronic structure at several energies, as indicated, recorded at $h\nu=60$ eV photon energy and $T=20$ K. Around $E=-50$ meV a Lifshitz transition from four small elliptical hole pockets to two large hole pockets occurs.
(e) - (h) Dispersion along the momentum lines marked in panel (f), parallel to $\bar\Gamma-\bar X$ at $k_c=0.05$, 0 and -0.05 \AA$^{-1}$. Thin dashed lines in (k) illustrate linear dispersion of the measured bands.}
\label{ARPES_S}
\end{center}
\end{figure}

At low energies, the in-plane electronic structure near $k_b=2n\pi/b$ seems to be more complicated than a simple model illustrated in Fig. \ref{ARPES}(e): it appears that near the FS, $E=0$, there might be four tiny isolated Fermi pockets, while the constant energy contours merge into two, further away from the Fermi level, as illustrated in Fig. \ref{ARPES_S}. 

The evolution of electronic structure with photon energy reflects the dispersion along $k_z$ ($k_b$ in our notation). Even though $k_z$ is not conserved in a photoemission process, it can be approximated by $k_{z}=\frac{1}{\hbar}\sqrt{2m(E_{k}cos^{2}(\theta)+V)}$, where $E_k$ is the kinetic energy of a photoelectron and $V$ is the inner potential. 
A set of ARPES spectra taken along $k_c=0$ in the SBZ at different photon  energies ($57\leq h\nu\leq 79$ eV, in 2 eV steps) is converted into a ($k_a, k_b, E$) data set by using the inner potential $V = 16$ eV. The $E=0$ slice, representing the FS is then shown in Fig. \ref{ARPES}(f). 
We note that in this photon energy range, the 10$^{th}$ and the 11$^{th}$ BZs are probed. 
States localized to the surface (surface states) do not disperse with $k_z$ and they would form straight vertical streaks in Fig. \ref{ARPES}(f), if they existed.

{\underline{Numerical estimates:}} 

To obtain a numerical estimate for CME, we take $T= 20 {\rm K}$, $\rho_{\rm Ohm} \simeq 1.2\ m\Omega\ cm$ indicated by our measurements at $B=0$, and assume that the rate of chirality-changing transitions $\tau_V^{-1} = \Delta/\hbar$, $\Delta \simeq 50$ meV and $\mu \sim 100$ meV as indicated by our ARPES measuments. The ARPES data also indicate that $v  \simeq 1/300\ c$. Evaluating Eq(\ref{cme_cond}), we reproduce the correct order of magnitide of the coefficient $a(T)$ that we used to fit the quadratic dependence of MR on magnetic field.

\end{document}